\documentclass[lettersize,journal]{IEEEtran}
\usepackage{amsmath,amsfonts}
\usepackage{algorithmic}
\usepackage{algorithm}
\usepackage{array}
\usepackage[caption=false,font=normalsize,labelfont=sf,textfont=sf]{subfig}
\usepackage{textcomp}
\usepackage{stfloats}
\usepackage{url}
\usepackage{verbatim}
\usepackage{graphicx}
\usepackage{cite}
\usepackage{graphicx}
\usepackage{amsmath}
\usepackage{array}
\usepackage{subfig}
\usepackage{multirow}
\usepackage{url}
\usepackage{booktabs} % For formal tables
\usepackage{adjustbox} % For adjusting table width
\usepackage{arydshln}   
\usepackage{multirow,multicol}
\usepackage{enumitem}
\usepackage{pifont}
\usepackage{amssymb} % For checkmark and xmark
\usepackage{wrapfig}
\usepackage{xcolor}
\usepackage{framed}
\colorlet{shadecolor}{gray!20}

\newcommand{\cmark}{\color{green}\ding{51}}%
\newcommand{\xmark}{\color{red}\ding{55}}%
\newcommand{\hlinewd}[1]{\noalign{\hrule height #1}}
\newcommand{\mfnt}[1]{\footnotesize{#1}}
\usepackage{rotating}

\begin{document}

\title{Towards Multimodal Emotional Support Conversation Systems}

\author{
        Yuqi~Chu,
        Lizi~Liao,
        Zhiyuan~Zhou,
        Chong-Wah~Ngo,~\IEEEmembership{Senior Member,~IEEE,}
        and Richang~Hong,~\IEEEmembership{Member,~IEEE}
        % <-this % stops a space
}

\maketitle

\begin{abstract}
The integration of conversational artificial intelligence (AI) into mental health care promises a new horizon for therapist-client interactions, aiming to closely emulate the depth and nuance of human conversations. Despite the potential, the current landscape of conversational AI is markedly limited by its reliance on single-modal data, constraining the systems' ability to empathize and provide effective emotional support. This limitation stems from a paucity of resources that encapsulate the multimodal nature of human communication essential for therapeutic counseling. To address this gap, we introduce the Multimodal Emotional Support Conversation (MESC) dataset, a first-of-its-kind resource enriched with comprehensive annotations across text, audio, and video modalities. This dataset captures the intricate interplay of user emotions, system strategies, system emotion, and system responses, setting a new precedent in the field. Leveraging the MESC dataset, we propose a general Sequential Multimodal Emotional Support framework (SMES) grounded in Therapeutic Skills Theory. Tailored for multimodal dialogue systems, the SMES framework incorporates an LLM-based reasoning model that sequentially generates user emotion recognition, system strategy prediction, system emotion prediction, and response generation. Our rigorous evaluations demonstrate that this framework significantly enhances the capability of AI systems to mimic therapist behaviors with heightened empathy and strategic responsiveness. By integrating multimodal data in this innovative manner, we bridge the critical gap between emotion recognition and emotional support, marking a significant advancement in conversational AI for mental health support. This work not only pushes the boundaries of AI's role in mental health care but also establishes a foundation for developing conversational agents that can provide more empathetic and effective emotional support.
\end{abstract}


\section{Introduction}
\begin{figure}[t]
  \centering
  \includegraphics[width=1.0\linewidth]{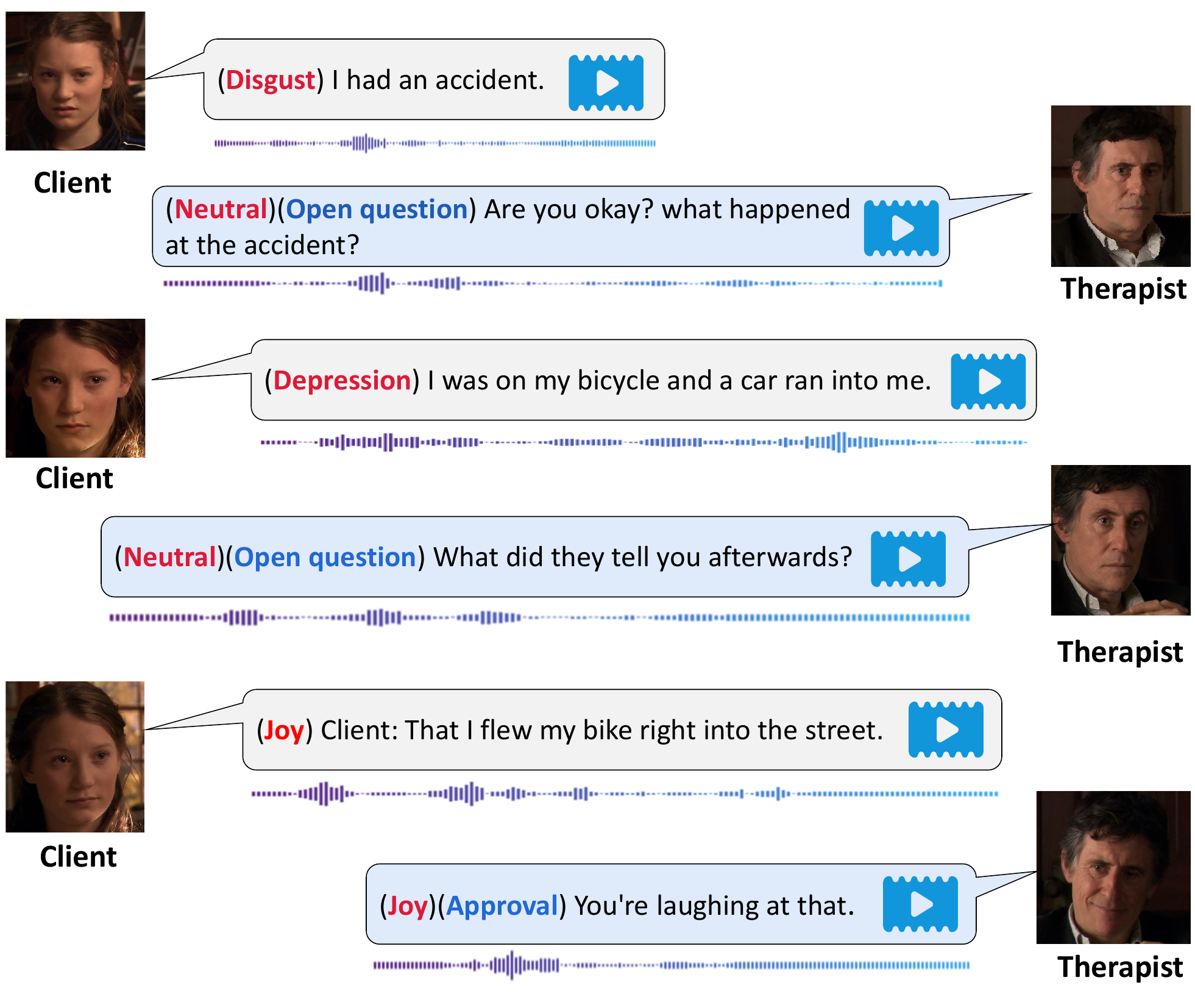}
  \caption{An example chat from the MESC dataset, where the client's and therapist's emotions are highlighted in bold red. Therapeutic strategies used by the therapist, informed by the client's emotions, are marked in blue, showcasing how multimodal information supports emotional engagement.}
  \label{fig1:example}
\end{figure}
%1. showing the importance of multimodal emotional support conversations.
The integration of conversational artificial intelligence (AI) into mental health care introduces a promising frontier for enhancing therapist-client interactions 
\cite{dingler2021use,jabir2023evaluating}. Aiming at replicating the rich nuances of human dialogue, conversational AI seeks to broaden the accessibility and depth of mental health support \cite{li2023systematic, torous2021growing}. This innovation stands to revolutionize the therapeutic landscape, offering the potential for more nuanced, empathetic interactions that bridge the gap between technology and the essential human elements of therapy, making effective mental health care more accessible to a wider audience.

\begin{table*}[!tp]
    % \vspace{-0.2cm}
    \caption{Comparison of all datasets for Emotion Recognition and Emotional Support Conversations. }
    \vspace{5pt}
    \label{tab:comparison}
    \centering
    \begin{adjustbox}{width=0.9\textwidth}
    \setlength{\tabcolsep}{1mm}{
    \begin{tabular}{lccccccc}
    \toprule
     & Text & Video& Audio & Emotion Recognition & Strategy Prediction & Response Generation\\
    \midrule
    Esconv \cite{DBLP:conf/acl/LiuZDSLYJH20}& \cmark& \xmark& \xmark & \xmark & \cmark & \cmark \\
    EmpatheticDialogues \cite{rashkin-etal-2019-towards} & \cmark& \xmark& \xmark & \cmark & \xmark  & \cmark \\
    DailyDialog \cite{DBLP:conf/ijcnlp/LiSSLCN17}& \cmark& \xmark& \xmark & \cmark & \xmark & \cmark \\
    EmotionLines \cite{DBLP:conf/lrec/HsuCKHK18}& \cmark& \xmark& \xmark & \cmark & \xmark & \xmark \\
   EmoryNLP \cite{zahiri2018emotion}& \cmark& \xmark& \xmark & \cmark & \xmark & \xmark \\
    MELD \cite{DBLP:conf/acl/PoriaHMNCM19} & \cmark& \cmark &  \cmark &  \cmark & \xmark & \xmark \\
    IEMOCAP \cite{busso2008iemocap} & \cmark & \cmark & \cmark & \cmark & \xmark & \xmark & \\
    \midrule
    \textbf{MESC} & \cmark & \cmark & \cmark &  \cmark& \cmark & \cmark \\
    \bottomrule
    \end{tabular}}
    \end{adjustbox}
% \vspace{-0.4cm}
\end{table*}

%2. discuss related works on existing methods.
Existing works has primarily focused on Emotion Recognition, using key multimodal benchmarks like the IEMOCAP \cite{busso2008iemocap} and MELD  \cite{DBLP:conf/acl/PoriaHMNCM19} datasets to recognize emotions in conversations. This area of study primarily identifies and tracks speakers' emotional states throughout a dialogue. For example, 
Li et al. \cite{li2023graphcfc} designed a Graph-based Cross-modal Feature Complementation (G-CFC) module to enhance modeling of contextual and interactive information across different modalities. 
Nie et al. \cite{nie2021gcn} developed an incremental graph convolution network (I-GCN) to capture both semantic correlations and temporal changes in utterances.
Ma et al. \cite{ma2023transformer} proposed a transformer-based model equipped with self-distillation (SDT) that effectively captures intra- and inter-modal interactions.
However, these efforts are limited to identifying emotional states and do not generate dialogues that respond to these emotions, thus failing to provide mental health support.
Furthermore, research on Emotional Support Conversation (ESC) has been solely based on text. Liu et al.  \cite{DBLP:conf/acl/LiuZDSLYJH20} introduced the ESC task along with the ESConv dataset to alleviate emotional distress through conversation. Tu et al. \cite{DBLP:conf/acl/TuLC0W022} and Peng et al. \cite{DBLP:conf/ijcai/00080XXSL22} advocated for the integration of commonsense knowledge into dialogue models to enhance their effectiveness. Cheng et al. \cite{DBLP:conf/acl/ChengS0CH23} developed the PAL method, which employs persona information and dynamically models conversation history to generate responses. Nonetheless, these approaches rely solely on text, overlooking other modalities and recognizing users' emotional states.

%3. pin-point two critical challenges: lack of dataset, lack of streamingline method. 
Despite these developments, to mimic the interactions between a client and therapist, particularly for addressing emotional distress, two critical challenges persist: 
\begin{itemize}

\item \textit{Challenge 1:}
The absence of a comprehensive multimodal dataset tailored for emotional support conversations. This gap significantly hinders the development of AI systems that can understand and respond to the complex emotional states of users  \cite{li2023systematic, torous2021growing}. 
As shown in Table \ref{tab:comparison}, existing datasets for emotional support conversations mainly concentrate on a single modality. For example, Esconv includes only text-based data, whereas multimodal datasets like MELD focus on identifying emotions in daily life dialogues. The datasets lack strategy and are more suited for emotion recognition than for generating therapeutic responses.

\item \textit{Challenge 2:} 
The lack of a streamlined methodological framework that integrates multimodal data for emotion recognition and generates empathetic and strategic responses in AI-driven therapy sessions \cite{acosta2022multimodal}. Existing methods typically treat emotion recognition, strategy formulation, and response generation as distinct, disjointed tasks. This approach fails to capture the interconnected nature of these components, which are considered in actual counseling conversations. Therapists need to consider the client’s emotions when generating responses and formulating treatment plans, which is crucial for empathy response and helpful for addressing emotion block.
\end{itemize}

%4. Introduce our construction of a new dataset, the significance, and how we construct and quality.
Addressing these challenges requires a comprehensive multimodal dataset and a general framework. 
(1) For \textit{Challenge 1},
we have constructed the MESC dataset\footnote{\url{https://github.com/chuyq/MESC}}. This first-of-its-kind dataset is enriched with comprehensive annotations, including emotions and strategies, across text, audio, and video modalities. As shown in Table \ref{tab:comparison}, the MESC dataset is versatile, supporting not only emotion recognition but also emotional support. These capabilities are essential for integrating conversational artificial intelligence (AI) into mental health applications.
(2) For \textit{Challenge 2},
We propose a general Sequential Multimodal Emotional Support Framework (SMES), a multi-task method grounded in Therapeutic Skills Theory \cite{anvari2020therapist, elliott1985helpful, hill2001effects, fosshage1997psychoanalysis}. The SMES framework leverages the strengths of multimodal foundation models to extract emotional cues from video and audio. It employs an LLM-based Reasoning model to sequentially generate multi-task results, encompassing user emotion recognition, strategy prediction, system emotion prediction, and response generation. By utilizing multi-task maximum likelihood training, the SMES framework adeptly models the interdependencies among these tasks, optimizing the therapeutic dialogue process in a comprehensive end-to-end manner.

To sum up, our main contributions are threefold:
\begin{itemize}
\item We introduce the first comprehensive multimodal conversation dataset for mental health care, combining text, audio, and video to capture the complex interplay of user emotions, agent strategies, and responses.

\item We develop a general
Sequential Multimodal Emotional Support Framework (SMES) based on Therapeutic Skills Theory, enabling AI to mimic therapist behavior more accurately through a sequential multi-task approach for emotion recognition and response generation.

\item We demonstrate significant improvements in AI’s empathy and strategic responsiveness for mental health support, establishing a new benchmark that bridges the gap between emotion recognition and emotional support.
\end{itemize}

\section{Related Work}
\subsection{Related Datasets for Emotional Support}
Emotional support has garnered attention due to its potential applications in psychology and emotional artificial intelligence systems. There are two mainly data-driven tasks: emotion recognition and emotional support conversations.
%introduce the emotion recogniton dataset and disadvantages
For emotion recognition, Li et al. \cite{DBLP:conf/ijcnlp/LiSSLCN17} developed the DailyDialog dataset, a text-based collection designed to mirror everyday communication styles and encompass various topics about daily life.
Chen et al.  \cite{DBLP:conf/lrec/HsuCKHK18} constructed the EmotionLines dataset, while Zahiri et al. \cite{zahiri2018emotion} annotated the EmoryNLP dataset. Both datasets are text-based and derived from the TV show \textit{Friends} with each utterance in these datasets annotated with one of seven emotion-categorical labels. The main distinction lies in the emotional classes assigned, and additionally, the EmotionLines dataset contains a larger number of utterances and dialogues compared to the EmoryNLP dataset.
Besides, Busso et al. \cite{busso2008iemocap} annotated the IEMOCAP database, a multimodal dataset comprising dyadic sessions in which actors engage in improvisations or scripted scenarios. This dataset features six emotion labels and encompasses a total of only 151 dialogues.
Poria et al. \cite{DBLP:conf/acl/PoriaHMNCM19} introduced the MELD dataset as an extension and enhancement of the EmotionLines dataset, which contains audio, visual, and textual modalities. MELD revisited the emotion labeling of the EmotionLines dataset, considering dynamic changes in emotional states observed in video data. While the EmotionLines dataset comprises 2000 dialogues larger than MELD contains 1400 dialogues. The inclusion of multiple modalities in MELD poses additional challenges for labeling, rendering the task more complex than text-only datasets.
% Summarize the shortcomings
While the existing datasets offer rich emotional labels, they lack emotional strategies. It only has the capable of analyzing the speaker's emotional state but can not provide corresponding emotional support tailored to that state.

%inrtoudece emotional support con
In the realm of emotional support conversations, Sharma et al \cite{DBLP:conf/emnlp/SharmaMAA20} annotated post-response pairs from TalkLife and mental health subreddits, with only the data from Reddit being publicly available. 
Hosseini et al. \cite{DBLP:conf/aaai/HosseiniC21} collected similar pairs from online support groups, although these dialogues are restricted to single-turn or brief interactions. 
Liu et al. \cite{DBLP:conf/acl/LiuZDSLYJH20} developed the ESConv dataset, comprising 1,053 dialogues from daily interactions and featuring eight types of support strategies. 
% Summarize the shortcomings
However, these datasets are limited as they are exclusively text-based, which inadequately captures the interaction in human counseling and diminishes the potential effectiveness of emotional support. Traditional therapists often utilize multimodal cues, such as changes in facial expressions and voice tone, which are absent in these datasets.

\subsection{Related tasks for Emotional Support}
% introdece related task and the interplay among the task is important.
Emotional conversation systems are comprised of key tasks such as emotion recognition in conversation (ERC), emotional conversation, and empathetic conversation.
% introduce the task work, emotion recognition
Emotion Recognition in Conversation (ERC) aims to automatically identify and track the emotional states of speakers in dialogues by leveraging multimodal cues such as facial expressions, vocal tonality, and gestures. Research in this area has primarily focused on three methodologies: commonsense reasoning \cite{ghosal-etal-2020-cosmic, li2021past}, attention-recurrent networks \cite{hu-etal-2021-dialoguecrn, hu-etal-2023-supervised}, and Graph Neural Network approaches \cite{zhang2023structure, ren2021lr}. A notable work is by Nie et al. \cite{nie2021gcn}, which used a dynamic graph structure to capture semantic correlations and temporal changes in utterances.
%shortcomings
These methods aim to enhance the accuracy of emotion recognition in conversations. However, they stop short of generating responses based on identified emotions.
% emotional conversation and empathetic conversation
Emotional and empathetic conversation tasks are designed to produce responses aligned with pre-specified emotional cues \cite{huang2020challenges, zhou2018emotional, yang2024exploiting}. Gao et al. \cite{gao2021improving} concentrated on detecting these cues within conversations to generate contextually and emotionally coherent responses. Furthering this approach, Sabour et al. \cite{sabour2022cem} have incorporated external commonsense knowledge to enhance the system's understanding of users' emotions. These methods collectively aim to comprehend and appropriately respond to users' emotional states.
% emotional support conversation
Besides, another key area is emotional support conversation (ESC), which seeks to offer emotional support through social interaction, not professional counseling. A significant contribution by Liu et al. \cite{DBLP:conf/acl/LiuZDSLYJH20} introduced the multi-turn ESC dataset. Building on this work, Peng et al. \cite{DBLP:conf/ijcai/00080XXSL22} implemented a graph-based method, while Deng et al. \cite{DBLP:conf/acl/DengZ0L23} further enhanced the approach by integrating knowledge for improved context comprehension and employing strategy predictions to steer the generation of responses.
% shortcomings
However, these methods only rely on textual data for generating emotional support and lack the ability to dynamically recognize emotions.
% combine task is important and lack dataset
The current limitation in achieving human-like interactions stems from the lack of dynamic interplay among tasks in emotional conversation systems. 
A key factor contributing to this issue is lacking of datasets featuring multimodality, emotional labeling, and strategy information.

\begin{table}[t]
\centering
\renewcommand*{\arraystretch}{1}
\caption{Statistics of MESC.}
\begin{tabular}{lccc}
\hline
\textbf{Category} & \textbf{Total} & \textbf{Therapist} & \textbf{Client} \\
\hline
dialogues & 1019 & - & - \\
\#Utterances & 28762 & 10326 & 18436 \\
Avg. length of dialogues & 28.2 & 10.1 & 18.1 \\
Avg. length of utterances & 32.9 & 32.3 & 33.2 \\
\#Emotion & 7 & 7 & 7  \\
\#Strategy & 10 & 10 & - \\
\#Scenarios & 15 & - & - \\
\hline
\end{tabular}
\label{tab:MESC}
\end{table}

\section{MESC Dataset}
Facial expressions, vocal tonality, and body language are crucial for analyzing a user's psychological state, enabling conversational systems to enhance their capacity to mimic human mental health support.
Current datasets for emotional support, however, are predominantly text-based and fail to capture these crucial cues. Furthermore, existing multimodal emotion datasets are generally focused on daily scenarios and do not incorporate therapeutic strategies. To fill this gap, we construct a multimodal emotional support dataset (MESC). 

\subsection{Data Construction}
The MESC dataset is derived from the TV show \textit{In Treatment}, specifically covering seasons 1 to 3. This series chronicles the weekly sessions of psychotherapist Paul Weston with his patients, as well as his own counseling with a therapist. The data source is highly professional and the case within has been analyzed by the Director of Clinical Psychology at the Shanghai Mental Health Center. Additionally, \textit{The New York Times} has praised the series for providing a compelling insight into the psychopathology of everyday life.
To ensure the dataset is adaptable for multiple emotional tasks, we have annotated each utterance with its corresponding emotion. Moreover, we  have annotated utterances spoken by the therapist with the strategy employed. Each dialogue is segmented into counseling scenarios, complete with detailed descriptions of each scenario, as shown in Fig. \ref{fig:scenarios}. More annotation details will be provided in the later subsection.

Given the complexity of this multimodal emotional support task, we have invested considerable effort to ensure the quality, effectiveness, and comprehensiveness of the dataset. Our efforts are concentrated on the following aspects:

(1) \textbf{Multimodal Unified Timestamps}: We synchronize the timestamps across all modalities to solve the discrepancies in timestamps between text and video modalities. After manual filtering of chatting segments that are irrelevant to emotion support counseling in the TV series, we extract the starting and ending points from videos and accurately correlate them with dialogue utterances and vocal tonality to ensure seamless alignment across different forms of data.

(2) \textbf{Annotation Quality Control}: To ensure the quality of the annotations, we have written a tutorial that includes definitions of the strategies, examples of emotion classification, and a three-hour training session for annotators. Additionally, each annotator must pass a preliminary test before beginning official annotations, and only those who pass are permitted to proceed with the annotating process.

(3) \textbf{Comprehensive Coverage of Emotional Support}: We have segmented the videos based on the client's experiences and scenarios. The dataset features a diverse range of 15 scenarios, 10 therapeutic strategies, and 7 emotion categories, providing a thorough overview of the emotional support process. This dataset can be applied to a range of tasks, including emotion recognition, strategy prediction, and response generation, offering
comprehensive data for emotional support in mental health
care.

\subsection{Dataset Annotation}
Our dataset focuses on multimodal emotional support. We annotate emotional states and emotional support strategies by taking into account the interplay among three modalities, considering facial expressions, vocal tonality, language, and gestures. This complex task requires a significant investment of time and labor.

To enhance labor efficiency and reduce costs, we employ a large model like GPT-3.5 for coarse-grained annotation, followed by manual fine-grained calibration. The overall annotation accuracy of GPT-3.5 is about 25\%, which is low. Therefore, we employ three graduate students specializing in emotional support research as annotators for fine-grained calibration. 
They undergo training in our labeling methodologies and must pass a rigorous annotation test.
The annotators are required to watch video clips to identify and calibrate three key elements: the client’s emotional state, the therapist’s emotional state, and the therapeutic strategy employed. Throughout this process, annotators consider not just the textual content but also the facial expressions, gestures, and vocal nuances presented by both clients and therapists.  Each piece of data is calibrated by two annotators to ensure consistency.
In cases of discrepancies between the annotators' assessments, a third annotator would review the video clip and decide on the most accurate interpretation.

% Highlight annotation effort divided into emotion and strategy
\textbf{Emotion Annotation}: we require the annotators to identify and calibrate the emotional states of both clients and therapists through detailed observation of video clips. 
During this process, annotators are instructed to consider not only the text content but also facial expressions, gestures, and vocal nuance cues presented by both clients and therapists. They need to annotate each utterance with emotional labels chosen from a predefined set of classes, encompassing seven emotions: anger, sadness, disgust, depression, fear, neutral, and joy, as detailed in Table \ref{tab:emotionMESC}.

\textbf{Strategy Annotation}: To teach the annotators to label emotional support strategies, we have written a tutorial that includes definitions of the strategies and a three-hour training session for annotators.
Drawing inspiration from the online emotional support platform \cite{baumel2015online}, we develop ten sub-tasks. These sub-tasks are designed to help annotators learn the definitions of the ten professional therapeutic support strategies. Each sub-task is structured around an example conversation excerpt, accompanied by a quiz question tailored to cement the annotator's understanding of each therapeutic strategy. This educational approach ensures that annotators are acquainted with theoretical concepts while watching the video.

\begin{table}[t]
	\centering
 \renewcommand*{\arraystretch}{1}
 	\caption{Emotion and Strategy distribution in MESC.}
	\small
	\resizebox{\linewidth}{!}{
	\begin{tabular}{cc|cccc}
		\hlinewd{1.5pt}
		&& \multicolumn{4}{c}{{MESC}} \\
		\cline{3-6}
		\multicolumn{2}{c|}{{Categories}} & Train & Val & Test & Total \\
    \hlinewd{0.8pt}
  	\parbox[t]{2mm}{\multirow{7}{*}{\rotatebox[origin=c]{90}{\scriptsize\textbf{{Emotion}}}}}&anger & 1964  & 273 & 422 & 2659 \\
		&sadness & 597  & 82 & 80 & 759 \\
		&disgust  & 1406  & 73 & 190  & 1669  \\
		&depression  & 4033  & 498 & 427  & 4958 \\
		&neutral & 13762  & 1665 & 1689  & 17116 \\
		&joy & 1150  & 112 & 88  & 1350 \\
		&fear & 214  & 11 & 26  & 251 \\
	%	non-neutral & 2017 & 214 & 541
		\hlinewd{0.8pt}
		\parbox[t]{2mm}{\multirow{10}{*}{\rotatebox[origin=c]{90}{\scriptsize\textbf{{Strategy}}}}}&Open questions& 1892 & 253 & 258& 2403 \\
		&Approval& 610 & 73 & 76 & 759 \\
		&Self-disclosure& 1052 & 104 &  126 & 1282 \\
            &Restatement& 1011 & 143 & 128 & 1282 \\
		&Interpretation& 2124 & 278 & 307 & 2709 \\
            &Advisement& 390 & 52 & 41 & 483 \\
            &Communication Skills& 645 & 43 & 63 & 751 \\
            &Structuring the therapy& 208 & 12 & 26 & 246 \\
		&Guiding the pace & 300 & 41 & 45 & 386 \\
            &Others& 20 & 2 & 3 & 25 \\
		\hlinewd{0.8pt}
	\end{tabular}
	}
	\label{tab:emotionMESC}
\end{table}

\subsection{Quality Control}
We employ a variety of methods to ensure that the videos and dialogues selected for our multimodal dataset are of high quality and tailored for emotional support conversations.

\textbf{Timestamp Alignment and Content Validity}: To ensure the alignment of timestamps across the three modalities and maintain the validity of the dialogue content, we first write a script to manually calibrate each episode, aligning the subtitles closely with the videos. We then utilize the transcription alignment tool Gentle to achieve precise timestamp alignment for each sentence. This tool automatically aligns the transcript text with the audio and extracts word-level timestamps for accuracy. Furthermore, to maintain content relevance, we remove segments unrelated to emotional counseling, such as interactions between the therapist and his family members, thus focusing the content solely on patient interactions.

\textbf{Annotation Correction}:
To ensure data quality, we implement a two-tier annotation strategy. Initially, GPT-3.5 is employed for coarse-grained labeling of emotional states and therapeutic strategies. This is followed by meticulous checks and calibrations by our annotators. This approach not only reduces costs and labor but also aims to minimize labeling bias, providing a more balanced and nuanced understanding of the data. For emotion annotation, we require concordance between the labels from two annotators and manually calibrate 20,133 utterances—over 70\% of the data initially annotated by GPT. After these manual calibrations, we achieve an emotion Fleiss kappa score of 0.57, compared to a score of 0.43 for the MELD dataset.  For strategy annotation, considering the complexity of the task and the requirement for at least two annotators to agree, over 83\% of the labels undergo manual calibration. After these calibrations, we achieve a strategy Fleiss kappa score of 0.69.

\section{Dataset Characteristics}
\subsection{Statistics}
Our multimodal dataset MESC comprises 28,762 utterances, 1,019 dialogues, and each utterance is annotated with emotion labels from seven categories and ten therapeutic strategies. Additionally, the dataset includes a corresponding set of video and audio clips, matched in quantity to the utterances, to provide a comprehensive multimodal resource, as detailed in Table \ref{tab:MESC}.
% the length is enough for ES
The dataset reveals an average dialogue length of 28.2 utterances, pointing out that effective Emotional Support (ES) needs a relatively long and multi-turn conversation.  It highlights that clients need to share personal experiences fully. Therapists, in turn, need the information to explore where the emotional wounds originate from, thereby enabling them to formulate and apply targeted therapeutic strategies to distress the clients' stress.

\begin{figure}[t]
    \centering
    \includegraphics[width=0.9\linewidth]{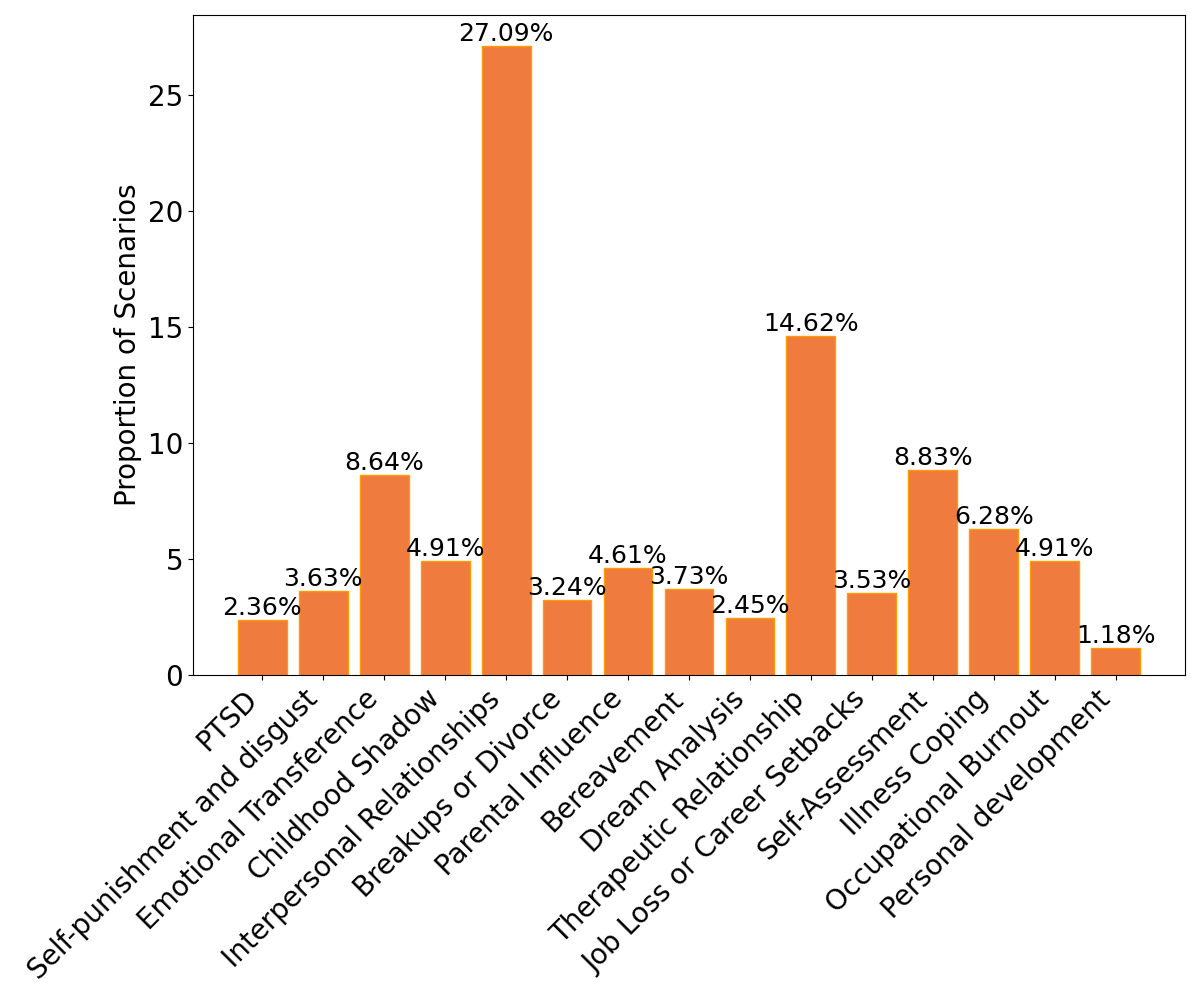}
    \caption{
    The proportion of scenarios of MESC. 
    }
    \label{fig:scenarios}
    % \vspace{-5mm}
\end{figure}

% anlysis the emotion and the reason from what experiencing
In our study, emotions are classified into seven distinct categories for annotation: anger, sadness, disgust, depression, neutral, joy, and fear. Different from the prior dataset MELD \cite{DBLP:conf/acl/PoriaHMNCM19} and IEMOCAP \cite{busso2008iemocap},  our dataset concentrates on the emotional states encountered in therapeutic counseling. The focus on dynamic emotion recognition is maintained throughout all stages of our research, including the training, validation, and testing phases.
% emotion distribution and scenarios
Our analysis of the dataset's emotional distribution uncover a non-uniform pattern, with neutral emotions emerging as the predominant category. This is attributed to therapists often keeping a neutral emotional state to create a trustworthy and communicative environment, encouraging clients to open up more freely. The second emotional state is depression, indicating that most of clients are experiencing emotional blocks, they are in a low mood coming from the relationship with friends and family, or a life-changing event, the distribution is shown in Fig. \ref{fig:scenarios}. 
% anlysis the proportion of scenarios 
It comprises 15 scenarios from clients face in life, including PTSD, dream analysis, childhood shadow, and other issues typically addressed in professional therapeutic settings. Among these, clients most frequently express concerns related to their familial relationships and their therapeutic interactions. This suggests that emotional blocks often originate from the following areas: clients may fear rejection or endure negative social interactions, prompting them to hide and suppress their pain. Therefore, providing effective emotional support requires creating a trusting and empathetic environment that helps clients express their feelings and thoughts, and understand their true emotions.

%METC has the richest annotation and can be used in various tasks
Beyond emotion annotation, we detail the statistics of strategy annotations in Table \ref{tab:emotionMESC}. The MESC dataset sets itself apart from existing datasets by not only focusing on emotion recognition but also on emotional support strategies, and emotional response generation. The MESC dataset is endowed with the most comprehensive annotations for emotional support tasks currently available. Consequently, it acts as a valuable resource for bolstering emotional support within AI conversational systems and can be applied to a wide array of emotional tasks.

\begin{figure}[t]
    \centering
    \includegraphics[width=1.0\linewidth]{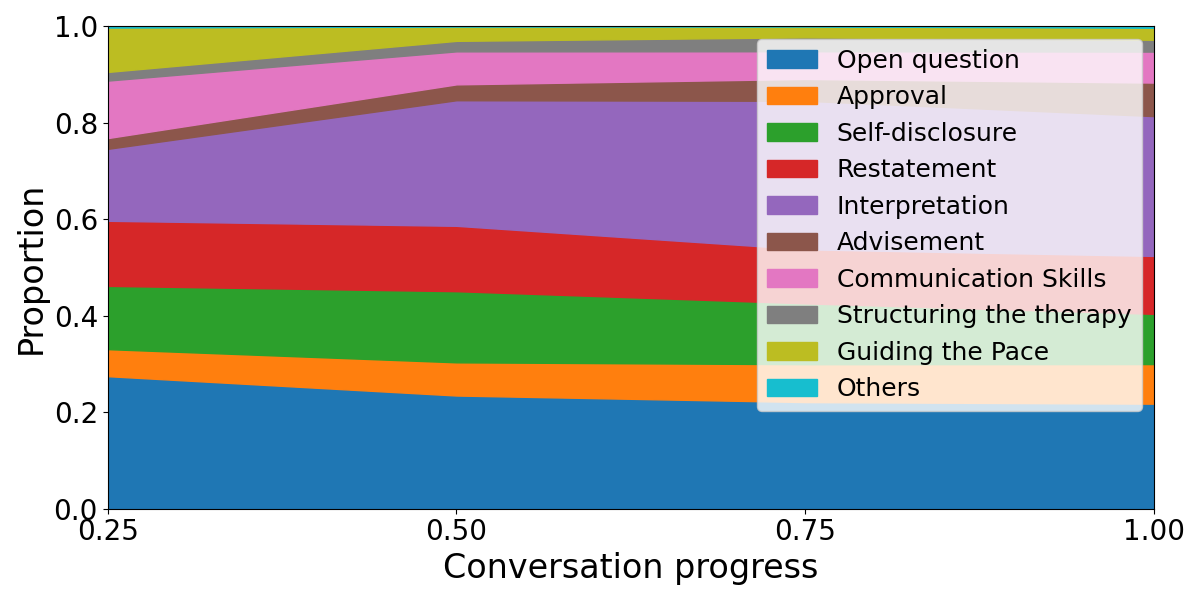}
    \caption{
    The distribution of strategies at different conversation progress.
    }
    \label{fig:distribution}

\end{figure}

\subsection{Strategy Analysis}
%strategy analysis
% strategy is related to emotional state
In our study, we aim to analyze the strategy employed by the therapist at different phases of emotional therapeutic counseling. To achieve this, we consider a conversation with $N$ utterances in total, where the $k$-th utterance from the therapist employs strategy $S$. The position of this utterance within the conversation is defined as the conversation phase and represented as $k/N$.
To visually display the changes in strategy employed during the dialogue process, we divide the progression of the conversation into four phases for analysis. Fig. \ref{fig:distribution} shows the distribution of ten strategies across the conversation progress, derived from professional therapeutic theories\cite{anvari2020therapist, elliott1985helpful, hill2001effects, fosshage1997psychoanalysis}.

It is noteworthy that the choice of strategy is influenced by changes in the client's emotional state. For instance, therapists might use ``open questions'' to explore underlying issues if the client displays a low mood, or ``approval" to provide positive reinforcement, thus fostering more open communication. Additionally, significant shifts in the client's mood during a session may prompt therapists to engage in ``self-disclosure" or ``guide the pace and depth of the conversation," aimed at managing the client’s emotional state and enhancing the therapeutic relationship.
Importantly, a therapist can employ multiple strategies within a single phase of progress. As depicted in Fig. \ref{fig:distribution}, the strategy of ``Open Questions" is consistently observed across all four stages, with a relatively high frequency and in combination with other strategies. In the initial stage, ``Open Questions" are paired with ``Restatement," enabling therapists to probe into the origins of the clients’ emotional distress. As the therapy progresses to the middle and later stages, this strategy is complemented by ``Interpretation", designed to help clients process their emotions, uncover the root causes of their issues, and reduce stress.

\section{Methodology}
\subsection{Task Definition}
\begin{figure*}[!tp]
  \centering
\includegraphics[width=1\linewidth]{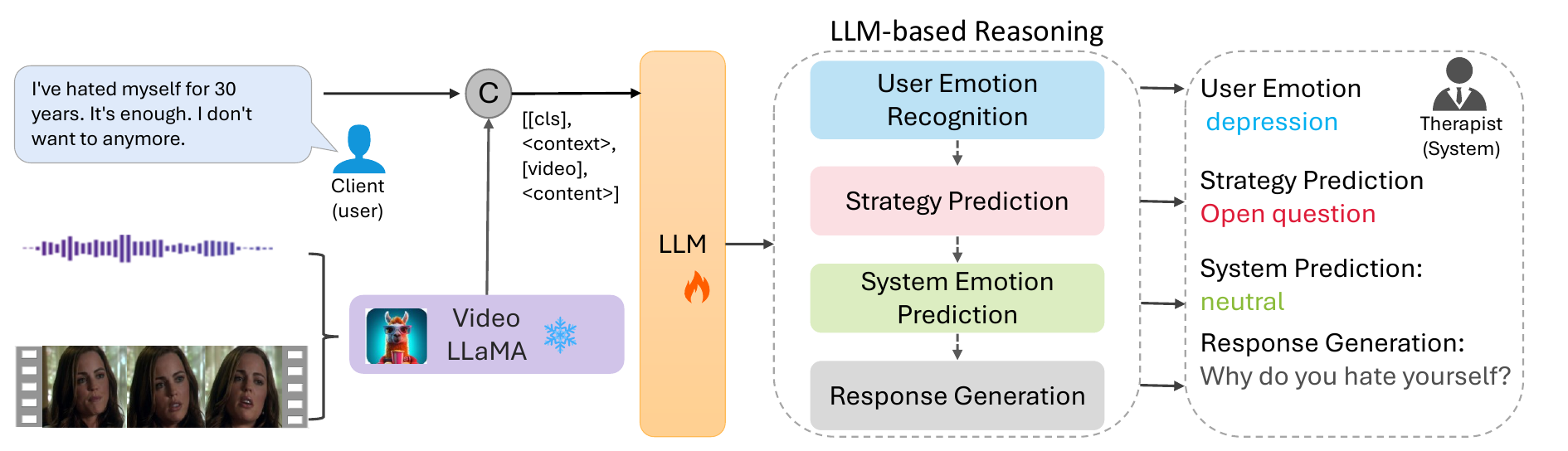}
\vspace{-0.5cm}
  \caption{The SMES framework uses multimodality information as inputs to improve mental health support. It employs Video-Llama to extract emotional cues from video and audio, then processes them through the LLM-based Reasoning model to sequentially generate four emotional-related task results.}
  \label{fig:framework}
\end{figure*}
To replicate the functions of a human therapist in providing emotional support, there are four critical tasks:

\vspace{+0.1cm}
\noindent \textbf{User Emotion Recognition}: This task involves identifying the emotional state of the client using multimodal inputs such as video, audio, and text. By analyzing facial expressions, body language, and voice intonations, alongside textual analysis of dialogue, the system dynamically models the client’s psychological condition. This comprehensive understanding is crucial for tailoring the conversation to the client's emotional needs.

\vspace{+0.1cm}
\noindent \textbf{System Strategy Prediction}: Based on the recognized emotions and the context of the conversation, the system predicts a therapeutic strategy. This involves choosing the most appropriate conversational approach, such as asking open questions, engaging in self-disclosure, or employing specific communication skills to address the client's underlying issues and alleviate stress. The strategy adapts to changes in the client’s mood and emotional state throughout the session.

\vspace{+0.1cm}
\noindent \textbf{System Emotion Prediction}: This task requires the system to predict its own emotional tone in responses, to align with the therapeutic strategy. By generating empathetic responses that reflect understanding and concern, the system fosters a supportive environment conducive to emotional healing.

\vspace{+0.1cm}
\noindent \textbf{System Response Generation}: The final task is generating responses that are not only contextually appropriate but also therapeutically effective. These responses are designed to resonate with the client's emotional state and therapeutic needs, helping to explore emotional wounds and promote psychological recovery. The dialogue generated by the system aims to support the client’s process of identifying and addressing emotional issues, contributing actively to their path toward emotional well-being.

\subsection{SMES Framework}

\begin{figure}[t]
    \centering
    \includegraphics[width=1.0\linewidth]{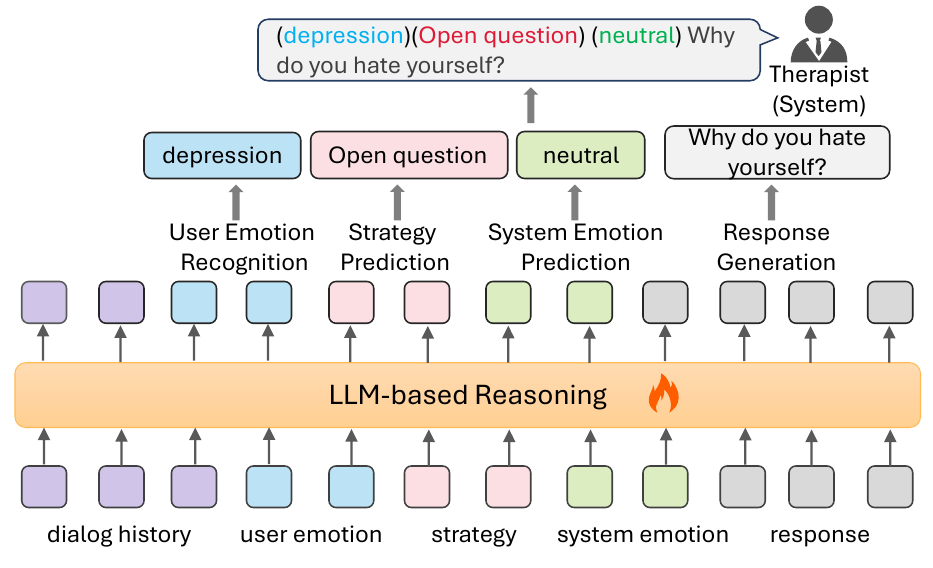}
    % \vspace{-3mm}
    \caption{
    The LLM-based reasoning modal consolidates all emotion-related sub-tasks into a sequence-to-sequence generation framework, optimizing them in an end-to-end manner.
    }
    \label{fig:training}
\end{figure}

%introduce the framework
In this section, we propose a general
sequential Multimodal Emotional Support framework (SMES)  designed specifically for emotional support. 
The framework is designed to leverage multimodal data to deliver rich emotional insights, supporting a range of tasks related to emotional support. It consists of four primary tasks: user emotion recognition, strategy prediction, system emotion prediction, and response generation. Illustrated in Fig. \ref{fig:framework}, during each turn $t$ of the dialogue process, the user provides inputs including an utterance $U_{t}$, video $V_{t}$, and audio $A_{t}$. After processing these multimodal inputs, they are transformed into a textual sequence $M_{t}$,  which   
captures the emotional state descriptions from this turn. The multimodal dialogue history can be represented as $H_{t}=\{M_{0}, R_{0}, \cdots, M_{t}\}$.
Our goal is to sequentially generate results for the four emotion-related tasks to provide better emotional support.

%introduce the modality transform
To achieve this, In a turn $t$,  we first employ an Audio-Visual Large Language Model (Video-LLaMA) \cite{zhang-etal-2023-video} to extract emotion-related cues
from video $V_{t}$ and audio $A_{t}$. we construct the prompt to query LLM as:
\begin{shaded}
{
\small
\noindent Video [$V_t$]; Audio [$A_t$]:\\
Question 1: ``What is the emotional state of the speaker?''\\
Question 2: ``What life distress might explain the speaker’s emotional expression and posture in this video?''
}
\end{shaded}

These questions enable the Video-LLaMA to detect emotional changes in visual scenes and audio signals, the clues are like 
"\textit{The speaker seems to be in a state of contemplation or thoughtfulness, as she is looking directly into the camera with a serious expression on her face.}"
The emotion clue can be denoted as $C_{t}$,  These are then concatenated with the user's utterance $U_{t}$ to form $M_{t}$.

\begin{equation}\label{eq:input}
M_t = [C_{t}, M_{t}] .
\end{equation}

To sequentially generate the four task results, the LLM-based Reasoning first reads all previous turns history $H_{t}$. It then generates user emotion recognition result $E_{t}$, 
\begin{equation}\label{eq:emotion}
E_t = \text{LLM-based Reasoning}(H_{t}) .
\end{equation}

Subsequently, the LLM-based Reasoning takes it as input to generate strategy prediction $S_{t}$. $S_{t}$ represents the strategy predicted by the system, which determines the therapeutic approach for generating responses—whether to inquire further about the situation or to offer sympathy and comfort. the LLM-based Reasoning then takes the concatenated sequence of $H_{t}$, $E_{t}$, and $S_{t}$ to decide the system emotion, $SE_{t}$ which influences the style of the responses generated by the system. The response $R_{t}$ is generated based on all prior information concatenated into a single sequence:
\begin{equation}\label{eq:response}
R_t = \text{LLM-based Reasoning}([ H_t,E_t,S_t,SE_t]) .
\end{equation}

Fig. \ref{fig:training} shows the training of the LLM-based Reasoning model. To capitalize on the strengths of pre-trained language models (PLMs) like BlenderBot, which has demonstrated a superior ability to generate high-quality responses in dialogue systems \cite{roller-etal-2021-recipes}, We integrate the generative PLMs into our framework and we reformulate a single training sequence as $Y = [ H_t, E_t, S_t, SE_t, R_t]$,
the model is trained to minimize the loss function $\mathcal{L}$ over the dataset $D$, where $I$ is the sequence length:
\begin{equation}\label{eq:loss}
\mathcal{L} = -\sum_{j=1}^{|D|} \sum_{i=1}^{I} \log P(Y_i \mid Y_{<i}) .
\end{equation}

\begin{table*}[]
    \centering
    \renewcommand*{\arraystretch}{1}
    \caption{presents a performance comparison across four tasks. While the SMES and GPT-3.5, GPT-4 are utilized for all tasks, other methods specialize in singular aspects, supporting either emotion recognition, strategy prediction, or response generation.}
    \scalebox{0.99}{
    \begin{tabular}{lcccccccccc} % Adjusted to 11 columns
    \toprule
     & \multicolumn{2}{c}{Emotion Recognition} & \multicolumn{2}{c}{Strategy Prediction} & \multicolumn{2}{c}{System Emotion Prediction} & \multicolumn{4}{c}{Response Generation} \\
     \cmidrule(lr){2-3} \cmidrule(lr){4-5} \cmidrule(lr){6-7} \cmidrule(lr){8-11} % Corrected the ranges
     Model & Acc$_{\uparrow}$ & W-F1$_{\uparrow}$ & Acc$_{\uparrow}$ & W-F1$_{\uparrow}$ & Acc$_{\uparrow}$ & W-F1$_{\uparrow}$ & B-2$_{\uparrow}$ & R-L$_{\uparrow}$ & BERTScore$_{\uparrow}$\\
     \midrule
    DialogueGCN & 46.27 & 50.61 & - & - & - & - & - & - & - \\
    MMGCN & 55.8 & 57.58 & - & - & - & - & - & - & - \\
    MMDFN & 58.13 & 55.86 & - & - & - & - & - & - & - \\
    \midrule
    Blenderbot-Joint & - & - & 48.0 & 46.1 & - & - & 4.85 & 15.25 & 85.5 \\
    BBMHR & - & - & - & - & - & - & 1.31 & 15.38 & 86.6 \\
    GPT3.5 & 33.5 & 33.8 & 19.9 & 17.6 & 17.4 & 27.6& 1.01 & 4.60 & 84.5\\
    GPT4 & 15.73 & 15.80 & 9.73 & 11.25 & 14.04 & 14.09 & 4.98 & 9.96 & 84.6\\
     \midrule
    \textbf{SMES} & \textbf{54.6} & \textbf{46.8} & \textbf{49.0} & \textbf{20.2} & \textbf{96.1} & \textbf{64.0} & \textbf{5.13}  & \textbf{15.42} & \textbf{86.8}\\
    \bottomrule
    \end{tabular}
    }
    % \vspace{-5pt}
    \label{tab:all_tasks}
\end{table*}

\section{Experiments}
SMES serves as a versatile framework for a variety of emotional conversational tasks, including emotion recognition, strategy prediction, and response generation. We have evaluated several methods related to these emotional tasks on the MESC dataset.

\subsection{Experimental Setups}
\noindent \textbf{Evaluation Metrics:}~  As for automatic evaluation, we utilize Accuracy and Weighted F1 as metrics for emotion recognition, strategy prediction, and system emotion prediction. In line with prior studies on response generation, our evaluation includes BLEU-n (B-2), ROUGE-L (R-L), and BERTScore to assess the quality of generated responses. These metrics collectively provide a comprehensive overview of model performance across different tasks.

\noindent \textbf{Baselines:}~ We evaluate methods across four tasks: emotion recognition, strategy prediction, system emotion prediction, and response generation. For emotion recognition, we assess DialogueGCN \cite{DBLP:conf/emnlp/GhosalMPCG19}, which utilizes Graph Convolutional Networks to enhance Emotion Recognition in Conversation; MMGCN \cite{DBLP:conf/acl/HuLZJ20}, employing a graph structure to capture both intra- and inter-modality features; and MMDFN \cite{DBLP:conf/icassp/HuHWJM22}, which leverages speaker features and integrates multimodal contexts while minimizing redundancy. For strategy prediction and response generation, we compare two methods: Blenderbot-Joint \cite{DBLP:conf/acl/LiuZDSLYJH20}, an open-domain agent with developed communication skills, and BBMHR (BlenderBot for Mental Health with Reasoning) \cite{DBLP:conf/acl/ZhangNM23}, which uses GPT-3 as an expert tailored for mental health and reasoning enhancement.
As a baseline for these four tasks, we utilize two method: GPT-3.5, a large language model known for its strong communication abilities, and GPT-4.0, the latest and advanced large language model.

\subsection{Overall Performance}
Table \ref{tab:all_tasks} presents the principal outcomes of our proposed method in comparison to the baseline models across four distinct tasks. Our method, the SMES framework, distinguishes itself by its versatility, demonstrating aptitude across all tasks. This contrasts with other models that specialize in specific areas. The SMES framework exhibits a robust performance, not only in emotion recognition and strategy prediction but also in system emotion prediction and response generation. We analyze the
results from four aspects:

\vspace{+0.1cm}
\noindent \textbf{Emotion Recognition}: The SMES achieves an accuracy of 54.6\% in emotion recognition, outperforming the specialized DialogueGCN model and closely following the state-of-the-art MMDFN. This slight deviation in performance can be attributed to the SMES’s expansive capabilities, which, unlike models solely concentrated on emotion recognition, are designed to excel across a spectrum of tasks. 
Diverging from methods focused exclusively on identifying emotions, the SMES adopts a comprehensive framework, with the dual ability to understand and engage with users.  It is engineered to understand and dynamically interact with users, thereby facilitating the generation of responses that are not only contextually appropriate but also emotionally resonant. For the SMES, identifying emotions is the critical first step towards its overarching objective: capturing the user's emotional state to deliver tailored and efficient emotional support.

\vspace{+0.15cm}
\noindent \textbf{Strategy Prediction}: The SMES distinguishes itself with a 49.0\% accuracy in strategy prediction, significantly outperforming GPT-3.5's 19.9\% and slightly besting Blenderbot-Joint's 48\%. This performance highlights the SMES's capability to handle strategic aspects of emotional support tasks.

\vspace{+0.15cm}
\noindent \textbf{System Emotion Prediction}: Previous methods often neglected the system's emotional state in response generation, a key factor in creating empathetic interactions. In contrast, the SMES framework effectively incorporates this aspect, achieving a remarkable 96.1\% accuracy in predicting system emotions—significantly surpassing GPT-3.5. This precision enables the SMES to generate more empathetic responses, providing enhanced support and comfort to users dealing with distressing issues.

\vspace{+0.15cm}
\noindent \textbf{Response Generation}: The SMES outperforms both BlenderBot-based methods and large language models methods across all generation metrics.
These results not only affirm the effectiveness of the SMES algorithm in delivering emotional support but also underscore the necessity of the entire framework for multitasking. By leveraging the inherent dependencies between tasks, SMES optimizes their interactions, thus producing responses that consider the user’s emotional state, strategic needs, and system-predicted emotions. This approach adeptly models the interdependencies among these elements, enhancing the therapeutic dialogue process in a comprehensive, end-to-end manner.

\begin{table}
    \centering
    \renewcommand*{\arraystretch}{1}
    \caption{Results of human evaluation.}
    \setlength{\tabcolsep}{1.6mm}{
    \begin{tabular}{lccccccccc}
    \toprule
       \multirow{2}{*}{SMES vs.}& \multicolumn{3}{c}{W/o ft}& \multicolumn{3}{c}{BlenderBot-Joint}\\
       \cmidrule(lr){2-4}\cmidrule(lr){5-7}
        & Win & Tie & Loss & Win & Tie & Loss  \\
       \midrule
        Flu.  & \textbf{48\%} & 7\% & 45\% & \textbf{51\%} & 17\% & 32\%\\
        Ide.  & \textbf{74}\% & 4\% & 22\% & \textbf{57\%} &9\% &34\% \\
        Com.  & \textbf{49\%} &12\% &39\% &\textbf{63\%} &15\% &22\% \\
        Sug.  & \textbf{59\%} &10\% &31\% &\textbf{54\%} &20\% &26\% \\
        Ove.  & \textbf{74\%} &9\% & 17\% & \textbf{70\%} &14\% &16\% \\
        \bottomrule
    \end{tabular}}
    \label{tab:human_eval}
\end{table}

\subsection{Human Evaluation}
Following previous studies \cite{DBLP:conf/acl/LiuZDSLYJH20,DBLP:conf/ijcai/00080XXSL22}, we conduct a human evaluation to compare the generated responses of two models across five dimensions: 
(1) Fluency: Which bot's responses are more fluent and easy to understand?
(2) Identification: Which bot more accurately explores your experiences and provides responses relevant to your problems?
(3) Comfort: Which bot's responses are more comforting?
(4) Suggestions: Which bot offers more helpful and empathetic suggestions for your problems?
(5) Overall: Which bot provides better emotional support for dealing with life's distressing problems?

We randomly select 100 dialogues from the MESC dataset and involve three human participants to determine the Win/Tie/Lose outcome for each dialogue comparison. We compare the responses generated by MESC with those from two other baselines, BlenderBot-Joint and BlenderBot (without fine-tuning on MESC), as detailed in Table \ref{tab:human_eval}. The results show that: (a) SMES demonstrates significant improvement in the identification metric, achieving a success rate of 74\% against BlenderBot and 57\% against BlenderBot-Joint. This
improvement is likely attributed to the emotional cues from multimodal data, which enhances the model's ability to accurately recognize the user’s state. (b) SMES performs better in the suggestion metric, outperforming both models by over 50\%. This may suggest that generating responses based on emotional recognition and strategic prediction effectively identifies user confusion, leading to more targeted advice. (c) SMES achieved a 70\% success rate in overall metrics, demonstrating its ability to offer better emotional support.  Furthermore, it may be beneficial to optimize the inherent interdependencies among the four tasks, which is essential for helping users accurately identify their issues and receive effective suggestions.

\begin{figure*}[t]
	\centering
 	\includegraphics[width=0.86\linewidth]{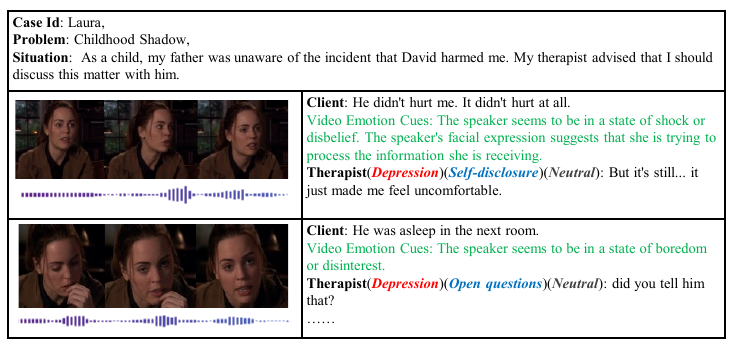}
	\caption{Case Study. 
    \textcolor[RGB]{0,176,80}{\textbf{Green text}}: The emotional cues are extracted from the video and audio.
    {\color{red}\textbf{Red text}}: The client's emotion is generated by the SMES method.
    \textcolor[RGB]{33,105,211}{\textbf{Blue text}}: The strategy is generated by the SMES method.
    \textcolor[RGB]{64,64,64}{\textbf{grey text}}: the therapist's emotion is generated by the SMES method.
    }
	\label{fig:case_study}
% 	  \vspace{-0.3cm}
\end{figure*}

\subsection{Ablation Study}
To evaluate the impact of each modality and sub-task on final performance, we conducted the ablation study, the results of which are presented in Tables \ref{tab:ablation0} and \ref{tab:ablation_emotion}. 

\noindent \textbf{Effect of multimodality information:} We first explore the impact of each modality's data on four tasks, as detailed in Table \ref{tab:ablation0}. Since we use video-llama for multimodality processing, where video and audio are bound together, removing video also means removing audio. 
We can observe that removing the text modality significantly impacts the four tasks, with emotional recognition dropping by 6.1\% and system emotion prediction by 25\%. Similarly, the video and audio modalities are crucial; their removal results in a 4.8\% decrease in strategy prediction and a 0.19 increase in the perplexity metric of generated responses. This demonstrates the importance of multimodal information in generating the four tasks and enhancing the effectiveness of emotional support.

\noindent \textbf{Effect of multi-task:} We conduct the impact of emotion and strategy tasks on response generation within an LLM-based reasoning framework. As detailed in Table \ref{tab:ablation_emotion},  removing the emotion task results in a 0.52 decrease in B-2 metric and a 1.27 reduction in R-L; Moreover, eliminating the strategy task has a greater impact on the response generation, leading to a 1.07 decrease in B-2 and a 1.30 reduction in R-L. This analysis reveals that a multi-task framework effectively harnesses the interconnected of these tasks to optimize response generation, making the responses more empathetic and supportive.

\begin{table}[!t]
    \centering
    \renewcommand*{\arraystretch}{1}
    \setlength{\tabcolsep}{3pt} % Set the column separation to 4pt
     \def\narrtablewidth{1.55 cm}
     \def\firsttablewidth{1.25 cm}
    \caption{Ablation studies for SMES, where `-text' and `-video' refer to the removal of the corresponding modality.
    }
    \begin{tabular}
    {
    p{\firsttablewidth} *{4}{p{\narrtablewidth}}
    }
     \hline
        \multirow{2}{*}{ Model }& \multicolumn{1}{l}{\textbf{Task1}} & \multicolumn{1}{l}{\textbf{Task2}} & \multicolumn{1}{l}{\textbf{Task3}} & \multicolumn{1}{l}{\textbf{Task4}}\\
        \cmidrule(lr){2-2} \cmidrule(lr){3-3}  \cmidrule(lr){4-4} \cmidrule(lr){5-5} 
        & Acc $_{\uparrow}$ & Acc $_{\uparrow}$ & W-F1 $_{\uparrow}$ & PPL$_{\downarrow}$ \\ 
        \hline
         SMES &54.6 & 49.0 & 64.0 & 14.18 \\
        \hdashline
        - video & 53.1\mfnt{($\downarrow$ 1.5)} & 44.2\mfnt{($\downarrow$ 4.8)} & 42.2\mfnt{($\downarrow$ 21.8)} & 14.37\mfnt{($\downarrow$ 0.19)} \\
        - text & 48.5\mfnt{($\downarrow$ 6.1)} & 46.5\mfnt{($\downarrow$ 2.5)} & 38.7\mfnt{($\downarrow$ 25.3)} & 16.82\mfnt{($\downarrow$ 2.64)} \\
        \hline
    \end{tabular}
    \label{tab:ablation0}
\end{table}

\begin{table}[!t]
    \centering
    \renewcommand*{\arraystretch}{1}
    \setlength{\tabcolsep}{3pt} % Set the column separation to 4pt
     \def\narrtablewidth{1.55 cm}
     \def\firsttablewidth{1.25 cm}
    \caption{Ablation studies for SMES, where `-emotion' and `-strategy' refer to the removal of the corresponding task.
    }
    \begin{tabular}
    {
    p{\firsttablewidth} *{4}{p{\narrtablewidth}}
    }
\hline 
Model & \multicolumn{1}{l}{\textbf{PPL}$_{\downarrow}$} & \multicolumn{1}{l}{\textbf{B-2}$_{\uparrow}$} & \multicolumn{1}{l}{\textbf{B-4}$_{\uparrow}$} & \multicolumn{1}{l}{\textbf{R-L}$_{\uparrow}$} \\
\hline
         SMES &14.18 & 5.13 & 1.37 & 15.42 \\
        \hdashline
        - emotion & 14.75\mfnt{($\downarrow$ 0.57)} & 4.61\mfnt{($\downarrow$ 0.52)} & 1.04\mfnt{($\downarrow$ 0.33)} & 14.15\mfnt{($\downarrow$ 1.27)} \\
        - strategy & 15.25\mfnt{($\downarrow$ 1.07)} & 4.06\mfnt{($\downarrow$ 1.07)} & 1.08\mfnt{($\downarrow$ 0.29)} & 14.12\mfnt{($\downarrow$ 1.30)} \\
        \hline
    \end{tabular}
    \label{tab:ablation_emotion}
\end{table}

\subsection{Case Study}
Fig. \ref{fig:case_study} presents a selection of dialogues from the MESC dataset along with responses generated by SMES, based on multimodal inputs. Initially, emotionally relevant clues are extracted from the video and audio inputs, as indicated by the green text in Fig. \ref{fig:case_study}. These modalities primarily focus on analyzing the speaker's emotional state and facial expressions. SMES then utilizes these clues, along with the user query, to perform four key tasks: recognizing user emotions, predicting strategies, forecasting system emotions, and generating responses. By integrating these four emotional tasks, the responses generated by SMES not only empathize with the user’s feelings but also effectively alleviate the user’s concerns, providing efficient emotional support.

\section{Conclusion}
In this work, we introduce the comprehensive multimodal MESC dataset for mental health care, along with the 
Sequential Multimodal Emotional Support Framework (SMES)—a general approach designed to enhance AI-driven conversation systems in mental health care. Developed using the MESC dataset and informed by Therapeutic Skills Theory, the SMES Framework skillfully extracts and integrates emotional cues from text, audio, and video modalities. By employing a sequential multi-task strategy that spans user emotion recognition, system strategy prediction, system emotion prediction, and response generation, this framework effectively captures the complex interplay of these elements to optimize therapeutic dialogues. 
Our MESC dataset and SMES Framework address two critical gaps: the lack of a comprehensive multimodal dataset for emotional support conversations and the absence of a cohesive framework for integrating multimodal data in conversational systems. 
Our extensive evaluation shows that the SMES significantly boosts the empathetic and strategic capabilities of AI, setting a new benchmark for conversational AI in mental health support.

\bibliographystyle{IEEEtran}
\bibliography{citation}
 
\end{document}